# Investigating Mixing Efficiency in Droplets: A Comprehensive Study of Numerical Modeling and Experimental Testing in 3D-Printed Microfluidic Devices

Ali Kheirkhah Barzoki[1,2], Alireza Mohseni[1,2], Mohammad Mehdi Bazyar[1,2], Kaivan Mohammadi[1,2,*]

[1] Department of Mechanical Engineering, Sharif University of Technology, Tehran, Iran

[2] Advanced Manufacturing Lab, Department of Mechanical Engineering, Sharif University of Technology, Tehran, Iran

Correspondence: kaivan.mohammadi@sharif.edu

## Abstract

Effective mixing is essential for biochemical reactions. In droplet-based microfluidics, immediate mixing of substances upon contact in the droplet formation stage can greatly enhance the uniformity of chemical reactions. Furthermore, it eliminates the need for implementing micromixers in the chip. In our research, we conducted a comprehensive study by first employing a series of two-dimensional (2D) numerical simulations, followed by experimental investigations using three-dimensional-printed (3D-printed) microfluidic chips. Our primary focus was on assessing the mixing efficiency within droplets. Specifically, we compared the performance of three different types of droplet generators: T-junction, cross-junction, and a novel asymmetric design with various angles. Our evaluation criteria encompassed mixing efficiency, droplet diameter, and droplet eccentricity. Our findings indicate that 2D numerical simulations can serve as a valuable tool for qualitatively analyzing two-phase flows and the droplet generation process, particularly in quasi-two-dimensional devices. The relative simplicity of such simulations renders them readily applicable, specifically in complex microfluidic geometries. Regarding mixing efficiency, we observed that the asymmetric droplet generators outperformed the cross-junction configuration but fell slightly short of the mixing efficiency achieved by the T-junction. Additionally, while the mixing index in the asymmetric generators closely matched that of the T-junction, these asymmetric generators produced smaller droplets. Our study suggests that the novel asymmetric droplet generators offer a significant advantage by simplifying the design of microfluidic devices. This is achieved by facilitating both droplet formation and rapid reagent mixing within the droplets while concurrently maintaining a small droplet diameter.

# 1. Introduction

Microfluidics is the fastest growing technology, and the microfluidics study is essential for implementing Lab-on-a-Chip. Lab-on-a-Chip systems are moreover recognized as micro total analysis systems ($\mu$TAS) capable of executing chemo- and bio-processes at the maximum possible scale (Lim et al., 2010). Biotechnology and biomedicine benefit greatly from microfluidic chips. Microfluidic devices find extensive application in chemical synthesis, drug delivery, cell separation, and solvent extraction owing to their notable advantages in enhancing mass transfer, reducing sample and reagent requirements, minimizing time and energy consumption (Battat et al., 2022; Kheirkhah Barzoki et al., 2023; Niculescu et al., 2021).

Among microfluidic systems, droplet-based microfluidic systems have captured considerable interest due to their capability to facilitate various functions, such as single cell analysis, chemical synthesis, polymerase chain reaction (PCR), loop-mediated isothermal amplification (LAMP), enzyme immobilization, and the production of specialized microparticles and nanoparticles (Abalde-Cela et al., 2018; Cedillo-Alcantar et al., 2019; Hattori et al., 2020; Li et al., 2022; Mao et al., 2019; Mbanjwa et al., 2018; Peng et al., 2020; Yang et al., 2021). The primary advantage of droplet-based microfluidics lies in its ability to precisely encapsulate extremely small volumes, ranging from femto- to nanoliters, of reaction components (Joanicot & Ajdari, 2005). This allows for rapid mixing and minimal thermal inertia within individual droplets, resulting in excellent control of reaction conditions. Furthermore, having droplets as the reaction chambers with low volume deviations, provides uniform conditions for the reaction. In all these applications, following encapsulation of biomolecules, single cells, or polymers with required chemical reagents in monodisperse water-in-oil or water-in-oil-in-water droplets, biochemical or chemical reactions are initiated. Various methods, such as fluorescent readers, are then used to detect the products of the reaction (Liu & Zhu, 2020; Payne et al., 2020). Although the performance of the microfluidic systems is highly dependent on efficient mixing in most of mentioned applications, the small size of microfluidic systems makes the flow in them laminar. Thus, mixing in these systems are mostly confined by diffusion making it difficult to achieve proper mixing (Bayareh et al., 2020; X. Chen et al., 2021).

Among various methods proposed to enhance the mixing process in microfluidic devices, one approach is to implement micromixers, involving both passive and active ones. Passive micromixers utilize specific geometries or microstructures within microfluidic channels to facilitate a more effective interaction between fluids (Bahrami & Bayareh, 2022; Lv et al., 2021; Shi, Huang, et al., 2021; Shi, Wang, et al., 2021; Wang et al., 2021; Yin et al., 2021). On the other hand, active micromixers employ control elements that require additional energy, such as electroosmotic, magnetic, and acoustofluidic mixing (Bahrami et al., 2022; Buglie & Tamrin, 2022; Z. Chen et al., 2022; Jalili et al., 2020; Lv & Chen, 2022; Mondal et al., 2021). Although achieving high mixing efficiency in passive mixers is challenging, active mixers enable more efficient mixing of reaction components at the cost of energy consumption and a more complex device design. In droplet-based microfluidic systems, several studies have been performed to increase the mixing index by adding micromixers to the region just after the droplet generator (C. Chen et al., 2018; Ghazimirsaeed et al., 2021; Harshe et al., 2016; Madadelahi & Shamloo, 2017; Yu et al., 2022; Yu & Chen, 2022). By this means, these curved structures induce advection in the droplets enhancing the mixing. To be noted that the implementation of micromixers adds complexity to the chip, increases its size, and by increasing the hydrodynamic resistance, limits the droplet generation frequency (Belousov et al., 2021). Consequently, there is a significant demand for alternative methods of mixing that do not rely on micromixers.

An alternative method to disrupt the flow involves the use of droplets. This approach involves the mixing of fluids during the formation of droplets. When two immiscible fluids come into contact at a junction, various forces such as viscosity, surface tension, and pressure gradient interact. As a result, the droplet eventually separates from the dispersed phase and continues to flow downstream. This method prevents the axial dispersion of mixing components by maintaining a restricted droplet size, allowing for rapid mixing through the internal circulation of the droplet (Burns & Ramshaw, 2001; Ward & Fan, 2015). However, Tice et al. showed that in a T-junction droplet generator, the effectiveness of this approach is highly dependent on the initial distribution of reagents. To address this sensitivity, one can adjust the relative flow rates of the continuous and dispersed phases (Tice et al., 2003). Belousov et al. designed an asymmetric flow focusing droplet generator enhancing the mixing in the droplet formation stage, and demonstrated that the mixing occurs six times faster than the symmetric one. It should be noted that flow focusing offers its own advantages. Its principle relies on the convergence of the continuous phase from two side channels with the dispersed phase at the intersection of the channels. At this intersection, the continuous phase compresses the dispersed phase, causing it to break up into droplets. This design allows for the prevention of droplets coming into contact with the walls of the channels, minimizing potential negative impacts on the components of the dispersed phase. Hence, it is logical to consider modifying the flow focusing design to enhance the mixing capabilities.

In this research, we performed a comprehensive study to compare the mixing efficiency, droplet diameter, and eccentricity of different droplet generator geometries. T-junction, cross-junction, and asymmetric flow-focusing junctions with various angles were analyzed. To model droplet formation and evaluate the fluid flows and distribution of reagents in the droplets, we conducted 2D numerical simulations. The simulation results were then compared to the experimental studies using 3D-printed chips. The simulations demonstrated that the design of the asymmetric droplet generator led to variations in fluid flows and reagent distribution within the newly formed droplets, resulting in improved mixing efficiency with droplet diameter smaller than that in the T-junction geometry. Subsequent experimental studies confirmed the simulation findings. This research provides a significant insight into selecting the best design for enhancing mixing capabilities in droplet-based systems.

## 2. Methods

### 2.1 Computational Method and Governing Equations

We describe the governing equations of the fluids using the well-known Navier-Stokes equations, with the stress tensor modified to accommodate the effects of surface tension. These governing equations are expressed in **Eqs. (1) and (2)** for our incompressible fluids in the laminar flow regime, as the maximum fluid flow rate was kept at 50 μL/min ($Re = \frac{\rho u w}{\mu} \approx 1$).

$$\nabla . \vec{u} = 0 \tag{1}$$

$$\rho \frac{\partial \vec{u}}{\partial t} + \rho(\vec{u} . \nabla)\vec{u} = -\nabla P + \mu \nabla^2 \vec{u} + \vec{F_s} \tag{2}$$

where $\vec{u}$ is the velocity vector ($m/s$), $\rho$ is the fluid density ($kg/m^3$), P is the pressure ($Pa$), $\mu$ is the dynamic viscosity ($Pa.s$), and $F_s$ is the surface tension force ($N/m^3$). The dominance of these forces over the surface tension was compared using capillary number (Ca) and Weber number (We) defined for each phase (i) as follows (Zhu & Wang, 2017):

$$Ca_i = \frac{(\mu u)_i}{\sigma} \tag{3}$$

$$We_i = \frac{(\rho u^2 L)_i}{\sigma} \tag{4}$$

where $L$ is the characteristics length (m), u is the velocity magnitude (m/s), $\mu$ is the dynamic viscosity ($Pa.s$), and $\sigma$ is the surface tension coefficient (N/m) between oil and water in this study.

In our two-phase system, the Cahn–Hilliard equation was implemented in order to track the fluid-fluid interface (Bhopalam et al., 2023; Jacqmin, 1999):

$$\frac{\partial \phi}{\partial t} + \vec{u} . \nabla \phi = \nabla . (\frac{\gamma \lambda}{\varepsilon^2} \nabla \Psi) \tag{5}$$

$$\Psi = -\nabla . (\varepsilon^2 \nabla \phi) + \phi(\phi^2 - 1) \tag{6}$$

In **Eqs. (5) and (6)**, $\phi$ is the phase variable to ensure a smooth transition across the phase interface. $\Psi$ is an auxiliary variable, and $\varepsilon$ represents the interfacial thickness, which was set to half of the maximum mesh size. In **Eq (5)**, $\gamma$ is the mobility parameter proportional to the interface thickness squared $\gamma = \chi \varepsilon^2$, where $\chi$ is the mobility tuning parameter, which was set to 1 in this study. $\lambda$ is the mixing energy density expressed as $\lambda = \frac{3\sigma}{\sqrt{8}\varepsilon}$. The surface tension force, expressed in **Eq. (2)**, could then be determined by multiplying the chemical potential (G) by the gradient of the phase field $F_s = G\nabla\phi$. The chemical potential is defined as follows (Kim, 2012):

$$G = \lambda(-\frac{\partial^2 \phi}{\partial t^2} + \frac{\phi(\phi^2 - 1)}{\varepsilon^2}) \tag{7}$$

The primary benefit of employing the phase-field method in modelling two-phase flows lies in its ability to compute the displacement of the contact line while adhering to a no-slip boundary condition for fluid velocity. Additionally, it mitigates pressure discontinuities at the corners and averts the creation of artificial vortices in the regions where channels intersect. Furthermore, as an interface capturing technique, it enables the precise resolution of droplet breakup.

To model the mixing process, the convective-diffusive mass transport equation was exploited:

$$\partial c / \partial t + \vec{u} . \nabla c = D\nabla^2 c \tag{8}$$

where c and D are the dye concentration ($mol/m^3$) and the diffusion coefficient ($m^2/s$), respectively. The mixing index (MI) was then calculated using the dye concentration in each image pixel (in experiments) or mesh cell (in simulations) by the following formula (Qian et al., 2019):

$$MI \, (\%) = \left( 1 - \frac{\sqrt{\frac{1}{N}\sum_{i=1}^{N}(c_i - \bar{c})^2}}{\bar{c}} \right) \times 100 \tag{9}$$

where N is the total number of pixels/cells within the droplet and $\bar{c}$ denotes the average dye concentration inside the droplet.

It's important to note that in this investigation, the properties of the dye solution were assumed to be identical to those of DI water, with a density of 1000 $kg/m^3$ and a dynamic viscosity of 1 mPa.s. Additionally, the simulations utilized properties for olive oil, setting its density at 917 $kg/m^3$ and dynamic viscosity at 84 mPa.s. The surface tension was established at 23.6 $mN/m$ for the interfacial tension between olive oil and water (FISHER et al., 1985). The boundary conditions employed in the simulations are presented in **Table 1**. V<sub>interface</sub>, P, and Q represent the velocity of the interface, pressure, and flow rate, respectively. As expressed in **Table 1**, a constant volumetric flow rate with a parabolic velocity profile for the fluid velocity was set for the inlet, and a constant gauge pressure was maintained for the outlet.

**Table 1:** Boundary conditions.

| | Boundary Condition |
|---|---|
| *Liquid-Solid Interface* | $u_{interface} = V_{interface} \ (no \ slip)$ |
| *inlets* | $Q = constant$ |
| *Outlet* | $P = 0$ |

It is assumed in these simulations that the two-phase fluids (i.e. water and oil) are (1) continuous, (2) incompressible, (3) isothermal, (4) immiscible, and (5) Newtonian with (6) constant physical properties in the (7) laminar regime using (8) two-dimensional geometries. Using finite element method (FEM), numerical modellings were conducted. To assess the mixing efficiency within the droplets, a system of three sets of equations needed to be simultaneously solved. The Navier-Stokes equations, phase-field equations, and the convective-diffusive mass transport equation are all solved simultaneously in the same computational framework. The convective-diffusive equation (Eq. 8) is solved in conjunction with the Navier-Stokes equations and the phase-field equations by considering the interactions between the fluid flow, phase interface, and mass transport process. The velocity field obtained from the Navier-Stokes equations influences the advection term in the convective-diffusive equation, while the phase field variable affects the diffusion and advection terms.

The fluid flow and phase-field equations were solved using the Parallel Direct Sparse Solver (PARDISO) with a residual tolerance set at 1E-3. Additionally, the mixing equation was solved using the Multifrontal Massively Parallel Sparse Direct Solver (MUMPS) with left preconditioning and a residual tolerance of 1E-3. Pressure discretization was accomplished using first-order elements, while velocity was discretized using second-order elements. Both phase-field and mixing solutions were discretized linearly. To linearize the set of non-linear equations at each time step, the Newton method was applied. The entire domain, including the boundary layers, was discretized using triangular (tri) elements. In this study, all simulations have been performed using 2D geometries. While 3D simulations provide the highest accuracy for the process, they demand substantial computing resources. However, 2D simulations still offer valuable insights and demonstrate satisfactory agreement with experimental data (Belousov et al., 2021).

## 2.2 Experimental Procedure

The schematic illustration of the experimental setup is depicted in **Fig. 1**. As shown in this figure, three insulin syringes (1 mL) were individually filled with olive oil (Merck), DI water, and a mixed solution of green food dye with DI water (1:10, w/w). In this study, the oil was used as the continuous phase, whereas DI water and the mixed solution of dye were prepared as the discrete phase. Then these liquids were injected to the 3D-printed microfluidic chip using three syringe pumps (SP204 SHS, FNM Co., Iran) with the ability to adjust the flow rate within the desired ranges: 5-20 μL/min for discrete phases and 50 μL/min for the continuous phase. Once the system reached a steady state, the droplet formation was recorded using a high-resolution digital light microscope (AM73115MZTL, Dino-Lite). The video frames were extracted, and then the mixing efficiencies were determined for five generated droplets using image-processing techniques in MATLAB (Mathworks).

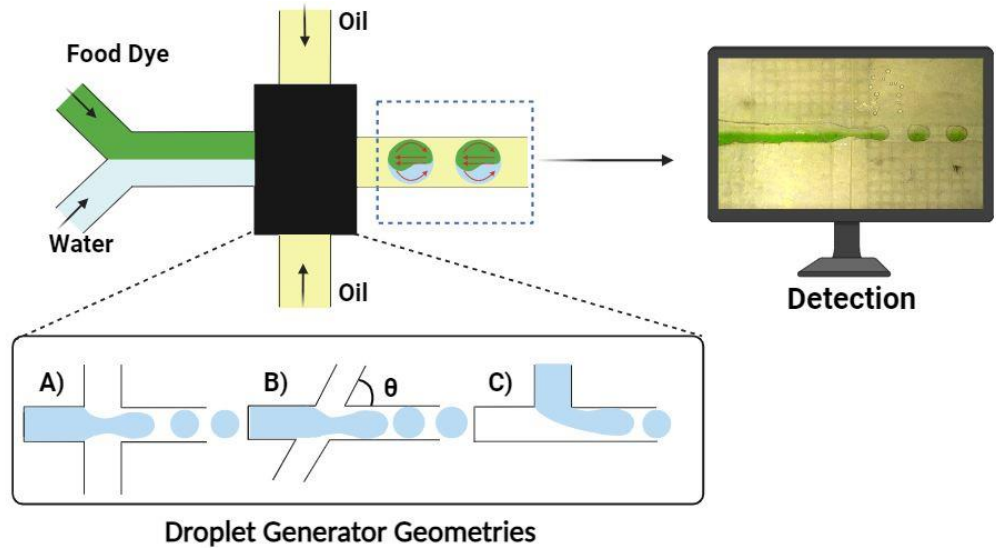

**Figure 1: The schematic representation of the setup containing three syringe pumps, a camera, a monitor, and microfluidic chips with different droplet generator geometries: A)** cross junction, **B)** asymmetric flow focusing, and **C)** T junction.

## 2.3   Microfluidic Chip Fabrication

We utilized a resin 3D printer to fabricate the chips, a method known for its simplicity and ability to produce high-quality chips with exceptionally smooth surfaces, devoid of any porosities within the channel walls. In contrast, similar studies have employed micromachining or soft lithography for chip fabrication. Chips created via micromachining exhibit surface roughness that does not match the level of smoothness achieved by resin 3D-printed chips (refer to the image below for comparison). On the other hand, chips produced through soft lithography contain porosity, specifically PDMS, within the channel walls, potentially impacting the mixing process. These limitations are absent when using a resin 3D printer, making it a superior choice for chip fabrication.

To fabricate the microfluidic chips, LCD 3D-printing method was utilized. A resin 3D-printer (Anycubic Photon mono X 6k) with a nominal resolution of 35 μm in the XY plane was used. The light emitting source of the printer was a string of 40 diode lights (405 nm, intensity: 22 mW/cm$^2$). Transparent plant-based UV resin (EC UV resin, Anycubic) was used as the printing material. The microchips were designed in CAD software (SolidWorks) and exported to a standard tessellation language (.stl) format for compatibility with the slicing software (Photon Workshop, Anycubic). Each microfluidic chip was sliced into 50 μm layers and printed at room temperature (25 °C). The exposure time was set to 25 s for the first 5 layers to ensure good attachment to the metallic platform, and then 1.1 s for the remaining layers. At the end of each layer, the UV light was turned off for 1 s to prevent unwanted parts from solidifying. After printing, the microfluidic chips were removed from the metallic platform and washed in a 99.5% solution of isopropyl alcohol (IPA) for 5 minutes to remove any unreacted resin residues. The microchips were then flushed with compressed air to evaporate any remaining IPA. The microchips were post-cured in a curing box (Washing and Cure 2, Anycubic) for 2 minutes to improve their mechanical properties. Once the 3D fabrication of the microchips was fully completed, they were washed with ethanol 95% and deionized water, dried using compressed clean air. **Fig. 2** demonstrates the fabricated chips prior to bonding. Afterwards, chips were bonded to polymethylmethacrylate (PMMA) substrates using thin layers of double-sided adhesive tapes. Next, chips were pressed for one day under the temperature of 80 ℃ to reduce air bubbles stuck between two layers of the chips. To be noted that before bonding chips, three holes for inlets and one hole for the outlet with a diameter of 2.5 mm were drilled through the PMMA sheet. It is important to drill holes before bonding to prevent drilling debris from going into the microchannel. To better connect tubes to the chip without leaking, pneumatic fittings (3 mm) were screwed through the PMMA layer.

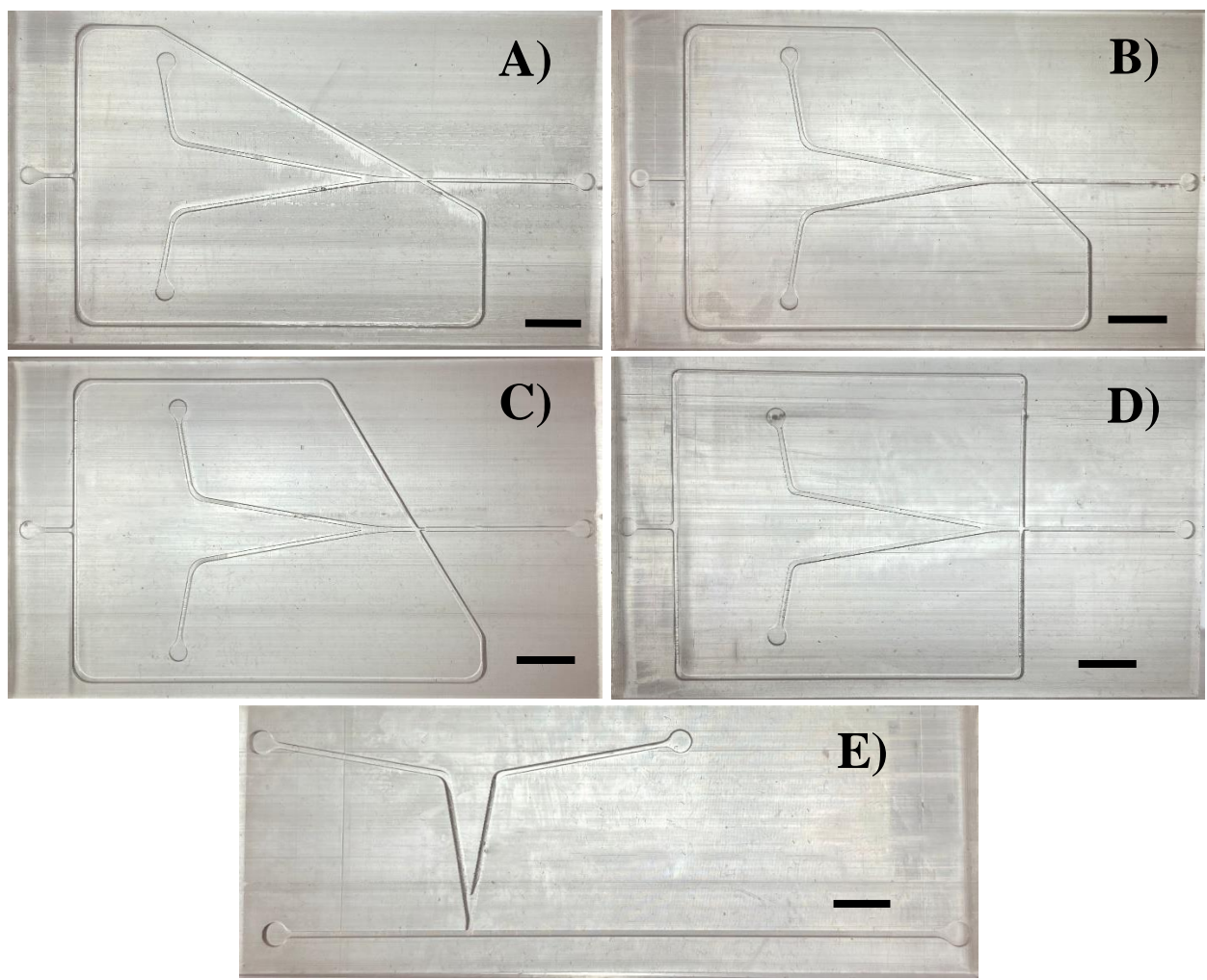

**Figure 2. 3D-printed microfluidic chips. A)** asymmetric 30°, **B)** asymmetric 45°, **C)** asymmetric 60°, **D)** cross-junction, and **E)** T-junction droplet generator. The error bars show 5 mm.

It is important to highlight that employing food dye to investigate the effectiveness of mixing within droplets and Dino-Lite microscope for video capture, along with the fabrication of chips using a resin 3D printer, established an alternative and uncomplicated setup. This stands in contrast to expensive setups involving PIV and soft lithography, yet it facilitated the acquisition of valuable findings.

## 2.4    Image Processing

First, a few image-preprocessing techniques, such as contrast enhancement and light adjustment, were applied to the original images extracted from the experiment results. Based on the pre-processed images, two approaches were implemented for the segmentation of the generated droplets: performing edge detection on the droplets using Sobel or Prewitt edge detectors and conducting image binarization with a variable threshold. In both scenarios, the Wiener filter was employed for noise reduction. Once the droplet segmentation was completed, the binary mask was convoluted with the original image to extract the frames associated with the inside of droplets, where the mixing occurs. The mixing index (MI) was then calculated using **Eq. (9)**.

# 3. Results and Discussion

## 3.1    Fluid Dynamics of Droplet Formation

As shown in Fig. 3, simulations of the droplet formation process indicate that, during the filling stage when two phases enter the junction before droplet formation, the velocity of the interface is notably lower than that of the continuous phase. This discrepancy results in the creation of either one or two fluid recirculation vortices within the dispersed phase. These vortices are induced by the flow of the continuous phase, primarily due to the boundary conditions at the liquid-liquid interface. In the case of the asymmetric design, this single vortex lacks symmetry with respect to the axis of the output channel. Consequently, it has the potential to impact the distribution of reagents inside the droplets and promote mixing. In contrast, in the cross-junction design, two symmetrical vortices exist, but they do not contribute significantly to the mixing process. In asymmetric geometries, there are two separate flows of the continuous phase around the dispersed phase (see **Fig. 3F**). However, due to the inherent asymmetry in these configurations, one of these flows, characterized by a smaller angle $\alpha$ in relation to the outlet axis, exerts a more substantial influence on the formation of the recirculation vortex within the dispersed phase when compared to the cross-junction geometry. Consequently, this dominant flow dictates the direction of the vortex. It's important to note that these two continuous phase flows act in opposite directions, leading to a partial attenuation of the dominant flow's impact on the vortex. In the T-junction geometry, similar to the asymmetric junctions, one single recirculation vortex is generated. However, in this scenario, during the filling stage, one side of the dispersed phase comes into contact with the continuous phase, which differs from the setup in the asymmetric geometries. Therefore, the preventive effect associated with the presence of a secondary continuous phase flow, as observed in asymmetric geometries, does not manifest in this situation. This lack of an inhibitory effect may result in a higher velocity of the vortex within the droplet.

In asymmetric designs, as depicted in **Fig. 3**, when we reduce the angle θ between the continuous phase inlets and the dispersed phase inlet, two notable effects occur: firstly, the size of the droplets during the filling stage increases, and secondly, the filling time also elongates. Consequently, the recirculation vortex within the droplet is afforded additional time for thorough mixing of the species inside. Moreover, as we reduce the angle θ, effectively narrowing the angle between the dominant continuous phase flow and the outlet axis, this decreases the attenuating effect of the weak continuous phase flow, leading to a heightened recirculation velocity within the droplet (see **Fig. 3F**). To be noted that in asymmetric generators, the vortex moves in the dispersed phase and this has a considerable effect in the mixing process (see **Videos S1-3**).

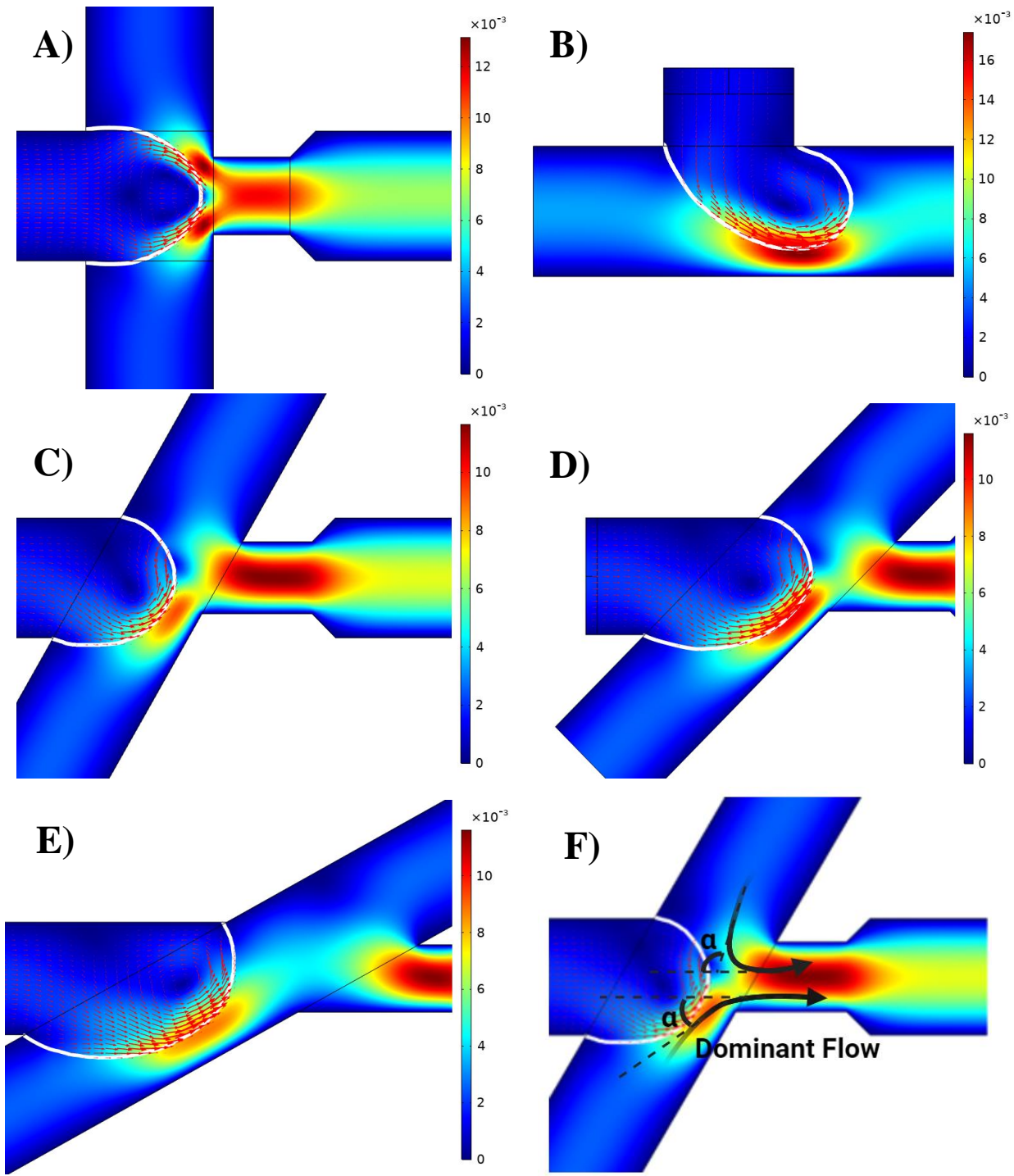

**Figure 3. 2D simulation of the droplet formation.** Velocity distribution during the filling stage in **A)** cross-junction, **B)** T-junction, **C)** asymmetric 60°, **D)** asymmetric 45°, and **E)** asymmetric 30°. **F)** Dominant flow in the asymmetric geometries and the angles of the continuous phase flows with the outlet axis ($\alpha$). The width of the inlets is 500 μm. The flow rates of the continuous and dispersed phase are 50 μl/min and 20 μl/min, respectively in all cases. The color bars show the velocity magnitude in m/s.

In most of the applications of droplet-based microfluidics, it is of utmost importance to isolate the formed droplets from being in contact with the channel walls. This prevents the droplets and the reagents inside them from being polluted. Having this in mind, producing small droplets would be highly sought-after. Consequently, along with mixing efficiency inside droplets, the size and eccentricity of the droplets is an important parameter in comparing the droplet-generator geometries. **Fig. 4** depicts the variations of droplet diameter and eccentricity as a function of the dispersed phase flow rate within various geometries. In this context, eccentricity is defined as $e = \sqrt{1 - b^2/a^2}$, where a and b denote the longest and the shortest diameters of the ellipse, respectively. This eccentricity value serves as a measure of how the droplet's shape deviates from a perfect circular form. Across all geometries, as the dispersed phase flow rate is raised, there is a consistent increase in the eccentricity of the droplets. This phenomenon occurs because the enlarged droplet size surpasses the channel's width capacity, resulting in deformation from its natural spherical shape. When comparing these geometries, it becomes evident that the T-junction geometry exhibits the highest values for both droplet size and eccentricity, while the smallest values are observed in the cross-junction geometry. In the context of asymmetric geometries, reducing the angle $\theta$ from 60° to 45° and 30° leads to a concurrent increase in both droplet diameter and eccentricity. This pattern can be attributed to the extended filling time, a consequence of the decreased shear force applied to the dispersed phase by the continuous phase, which in turn triggers the pinch-off of the droplet. When we reduce the angle $\theta$, the effective shear force diminishes, resulting in a delay in the pinch-off process.

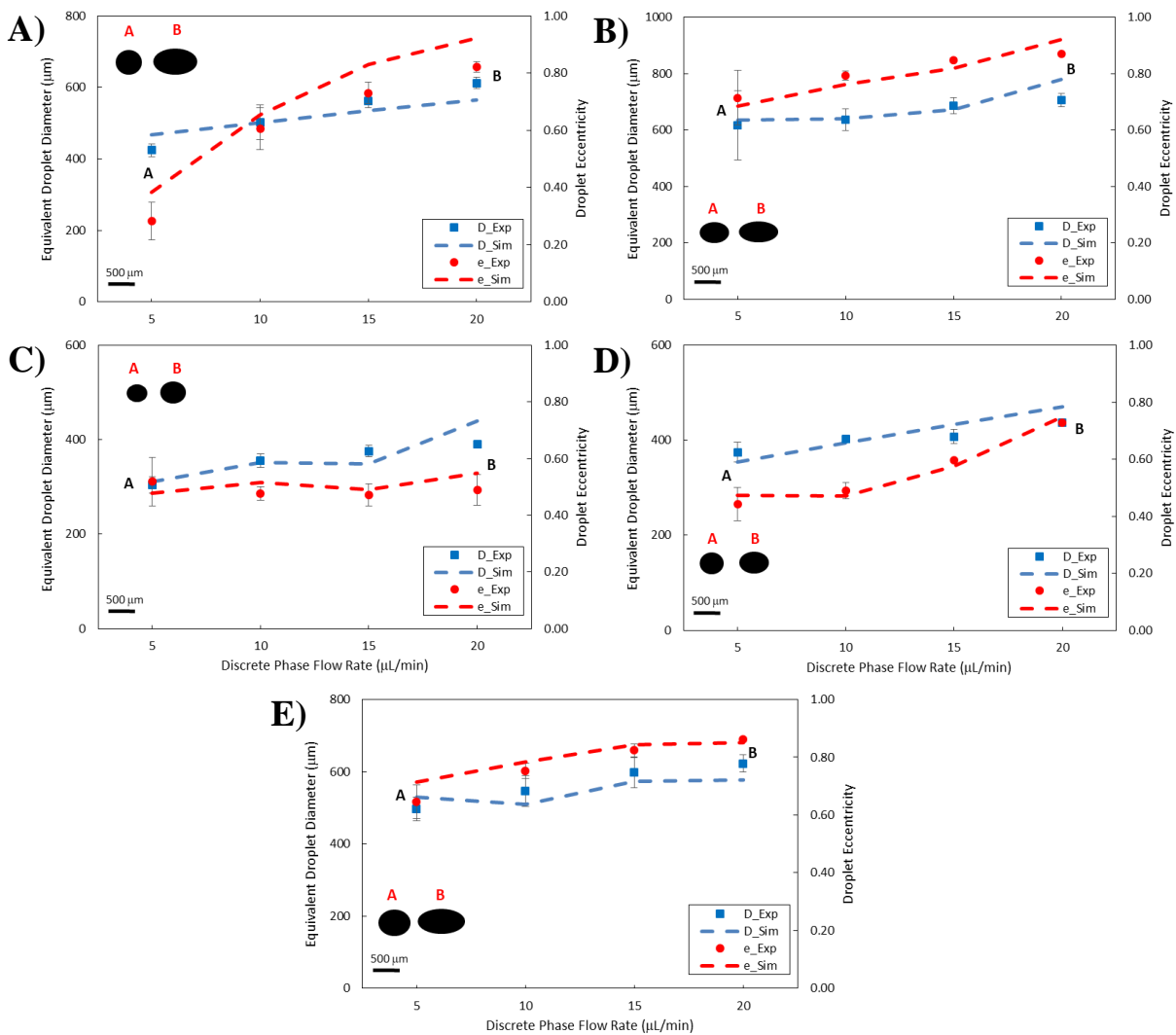

**Figure 4. Diameter and eccentricity variations of droplets with dispersed phase flow rate.** Comparative analysis of experimental results and simulations in various geometries: **A)** cross-junction, **B)** T-junction, **C)** asymmetric 60°, **D)** asymmetric 45°, and **E)** asymmetric 30°. Continuous phase flow rate is 50 µl/min.

## 3.2 Mixing Efficiency

Mixing plays a vital role in various applications, and the use of micromixers within microfluidic systems is a widely adopted technique to facilitate the proper mixing of reagents within droplets. However, the incorporation of micromixers can demand additional space. Furthermore, the immediate mixing of reagents upon contact can significantly improve the uniformity of chemical reactions and the products of the reaction. Considering all of these, the adoption of a droplet generator with the ability to mix reagents within the droplet during the filling stage appears to be a promising alternative to the use of micromixers. **Fig. 5** illustrates the assessment of mixing efficiency within droplets immediately after their formation using various droplet generators. It is worth noting that the mechanism of droplet formation and the characteristics of recirculation vortices inside the droplets significantly impact the mixing index. Across all geometries, there is a consistent trend: as the flow rate of the dispersed phase increases, the mixing index decreases. This phenomenon can be attributed to the reduction in filling stage duration, which occurs when the dispersed phase flow rate is increased. This implies that the species inside the droplet do not have sufficient time for thorough mixing due to the presence of formed vortices. Notably, the T-junction geometry displays the highest mixing index, which can be explained by the formation of a single large vortex within the droplet during the filling stage (**Fig. 3B, Videos S5,10, V5**). In contrast to other geometries, during this stage, only one side of the droplet is in contact with the continuous phase, resulting in a vortex with greater velocity and consequently, rapid mixing. Conversely, within the asymmetric geometries, two separate flows of the continuous phase surround the droplet. These flows exert varying effects on the vortex inside the droplet, with the one possessing a lower angle $\alpha$ in relation to the outlet determining the vortex's direction. However, these two flows operate in opposing directions, and the weaker of the two diminishes the effect of the prevalent one, and consequently, the velocity of the vortex within the droplet (**Fig. 3C-E, Videos S1-3,6-8, V1-3**). As a consequence, the vortex is weaker compared to that in the T-junction geometry, resulting in a reduction in the mixing index. In the cross-junction geometry, there are two flows of continuous phase around the droplet. Due to the symmetry in this geometry, these two flows are identical. As a result, two identical vortices but in the opposite direction are formed inside the droplet. Therefore, in this geometry, the mechanism of mixing is completely different from other ones (**Fig. 3A, Videos S4,9, V4**). Here, the presence of two vortices results in the independent mixing of species within the two halves of the droplet. Because there is no single vortex enveloping the entire droplet, this unique configuration leads to the lowest mixing index values among all the geometries.

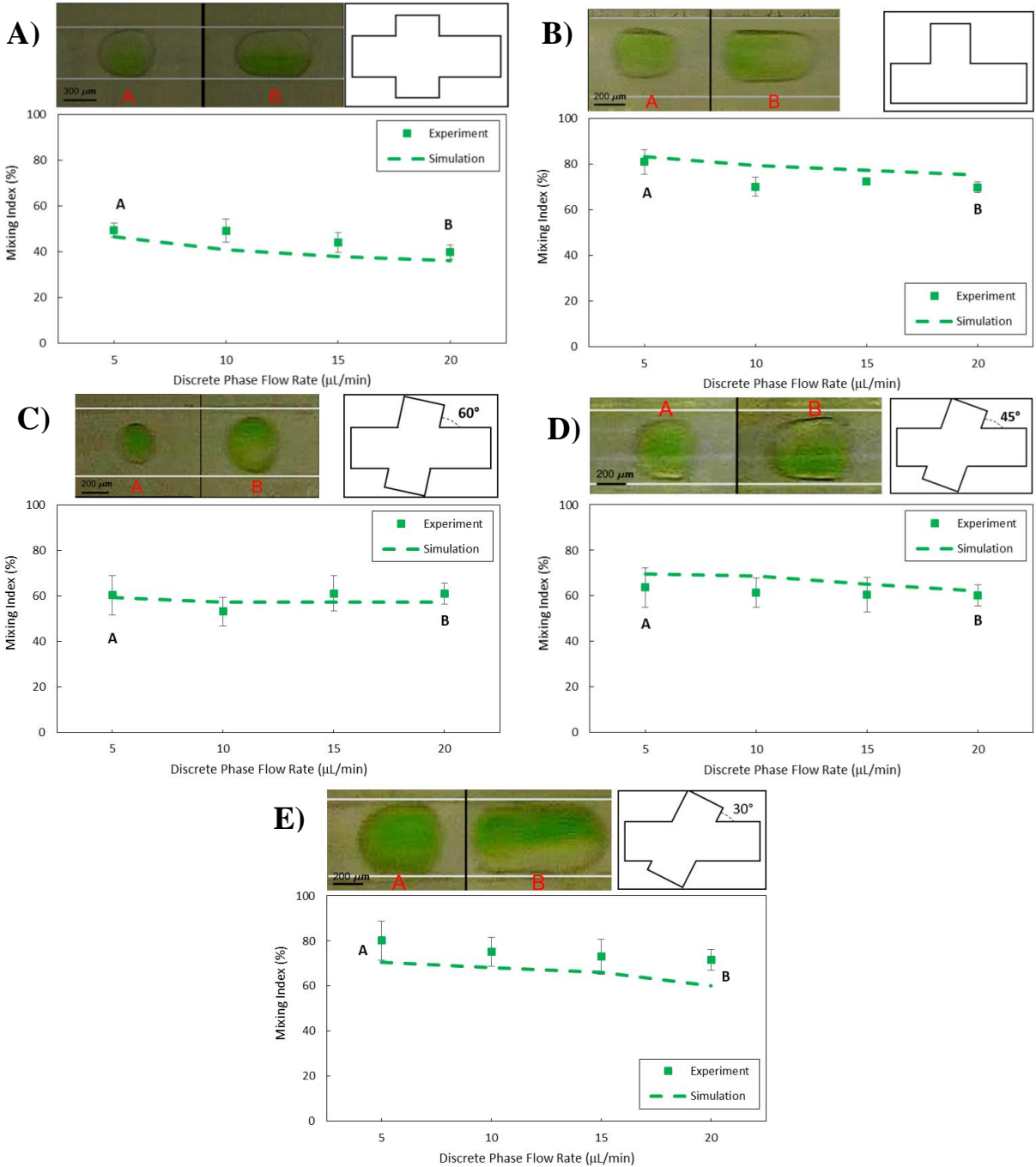

**Figure 5. Mixing efficiency variations within droplets with dispersed phase flow rate in different geometries**. Comparative analysis of experimental results and simulations in various Geometries: **A)** cross-junction, **B)** T-junction, **C)** asymmetric 60°, **D)** asymmetric 45°, and **E)** asymmetric 30°. Continuous phase flow rate is 50 μl/min.

Studying different angles of asymmetric designs reveals that decreasing the angle $\theta$ between the inlets of the continuous and dispersed phases increases the mixing index. Specifically, the asymmetric design with a 30° angle demonstrates the highest mixing index among all asymmetric designs. This mixing index is comparable to that of the T junction, while the droplet diameter is approximately 100 μm smaller than that in the T junction. Furthermore, as asymmetric designs implement flow-focusing droplet generation, the reduced contact of droplets with the channel walls decreases the likelihood of species contamination inside the droplets.

## 4. Conclusions

In our study, we undertook numerical simulation and experimental analysis of various droplet generator geometries. These included the T-junction, cross-junction, and asymmetric flow focusing droplet generators featuring distinct angles. We assessed parameters such as droplet diameter, eccentricity, and the mixing index within the formed droplets. One noteworthy observation pertained to the asymmetric designs, where a single asymmetric recirculation vortex was induced within the dispersed phase during the droplet formation stage. This phenomenon led to an enhancement in the mixing index in comparison to the cross-junction. The T-junction geometry, similar to the asymmetric counterparts, produced a single recirculation vortex, resulting in the highest mixing index among all the investigated geometries. The key difference between the T-junction and asymmetric geometries lies in the fact that the latter featured two continuous phase flows around the dispersed phase during the filling stage. These two flows somewhat attenuated their impact on the velocity of the vortex within the dispersed phase. However, in the T-junction, a single continuous phase flow led to a higher vortex velocity and, consequently, a superior mixing index. Conversely, the cross-junction geometry displayed the lowest mixing index across all flow rate ratios. This is attributed to the presence of two distinct vortices within the dispersed phase during the filling stage. These vortices independently mix the species in the upper and lower halves of the dispersed phase, precluding uniform mixing throughout the entire droplet. In the asymmetric geometries, we observed that reducing the angle between the inlets of the two phases correlated with an increase in the mixing index. It is noteworthy that the trend of mixing index variation, decreasing as the dispersed phase flow rate increased, was consistent across all the geometries studied.

Regarding the droplet diameter and eccentricity, the lowest and largest values were associated to the cross-junction and T-junction, respectively. This observation can be explained by the heightened shear force in the cross-junction, attributed to the presence of two perpendicular continuous phase flows intersecting with the dispersed phase flow. In the asymmetric geometries, a noteworthy trend emerged: as the angle between the phase inlets decreased, both the droplet diameter and eccentricity increased. This phenomenon stemmed from the reduced shear force associated with the decreased angle, resulting in extended filling times and the formation of larger droplets. Additionally, across all geometries, an increase in the dispersed phase flow rate corresponded to larger droplet diameters and eccentricities.

It is worth noting that, while the T-junction consistently exhibited the highest mixing index values across all flow rate ratios, it also produced the largest droplets. In contrast, the asymmetric geometries demonstrated comparable mixing index values to the T-junction but yielded smaller droplets. Consequently, the asymmetric geometries prove to be a favorable choice in applications where both high mixing efficiency and smaller droplet sizes are advantageous.

## Conflict of Interest

The authors declare no conflict of interest.